\documentstyle[12pt]{article}
\advance\textheight by \topskip
\begin{document}
\begin{flushright}
SU-ITP-95-33\\
HUB-EP-95/35\\
hep-th/9601010\\
\today\\
\end{flushright}
\vspace{1cm}
\begin{center}
\baselineskip=16pt

{\large\bf   F\,\&\,H~   MONOPOLES \, }  \\

\vskip 2cm

{\bf Klaus Behrndt}\footnote{E-mail:
behrndt@qft2.physik.hu-berlin.de}

Institut f\"ur Physik,
Humboldt-Universit\"at, 10115 Berlin, Germany \\
\vskip 0.6 cm
{{\bf Renata
Kallosh}}\footnote { E-mail:
kallosh@physics.stanford.edu}\\
 \vskip 0.2cm
Physics Department, Stanford University, Stanford   CA 94305-4060, USA\\
\vskip .6cm

\vskip 0.6 cm

\end{center}
\vskip 1 cm
\centerline{\bf ABSTRACT}
\begin{quotation}

  Supersymmetric monopoles of the heterotic string theory associated
  with arbitrary non-negative number of the left moving modes of the  string
states are  presented.  They include H
  monopoles and their T dual partners F monopoles (ALE instantons).
  Massive F = H monopoles are T self-dual. Solutions
  include also an infinite tower of generic T duality covariant
  non-singular in stringy frame F\&H monopoles  with the
  bottomless throat geometry.
The massless F = --H
  monopoles are invariant under combined T duality and charge
  conjugation converting a monopole into anti monopole.

  All F\&H monopoles can be promoted to the exact supersymmetric
  solutions of the heterotic string theory since the holonomy group is
  the compact $SO(9)$.  The sigma models for ${\bf M}^8$ monopoles,
  which admit constant complex structures, have   enhanced   world-sheet
  supersymmetry: (4,1) in general  and (4,4) for the left-right
  symmetric  monopoles. The space-time supersymmetric GS
  light-cone action in monopole background is directly convertible into
  the world-sheet supersymmetric NSR action.

\end{quotation}

\newpage

\section{Introduction}

Supersymmetric gravity has been  studied extensively over some number 
of years.
The properties of the electrically charged supersymmetric solutions have been
compared with the properties of the states
in string theories \cite{Sen}, \cite{DuffRum}. The results of such comparisons
indicate that many of the electrically charged solutions have the
interpretation as the string states. Much less is known about the magnetically
charged solutions. From the point of view of the low-energy effective actions
of
supergravity theories,  electric as well as magnetic and dyon configurations
come out as  solutions of semi-classical non-linear
equations. None of these soliton-type solutions
is directly and  unambiguously related to a  linear system of  excitations
describing the quantum states of the string theory.
However, the supersymmetric string theories and electrically charged solutions
seem to have a particular knowledge about each other. The predictions 
about the
properties of BPS string states sometimes were obtained in the framework of
soliton solutions and sometimes vice-versa. One of  the most striking examples
of such predictions was the one from the string theory. The massless states
with $N_L=0$ (where  $N_L$ is the number of left moving modes) were 
expected to
describe the T-self-dual point of the theory \cite{HullT}. Indeed, the
corresponding ``solitons" with the vanishing ADM mass were found to be a
T-self-dual solutions of the supergravity theory \cite{KB}. In addition, they
were identified with   $N_L=0$ states of the heterotic string theory 
\cite{KB}, \cite{K} \cite{KL1}.

The massless magnetic monopoles are also very interesting solutions, however
not much is known about them since the magnetic solutions have not been
identified directly with  excitations of any kind of a linear system. The best
information comes from the conjectured S-duality which tell us that S-dual
partners of massless electric solutions exist. The asymptotic form of magnetic
solutions of the heterotic string with $N_L\geq 1$ was found in \cite{Sen}.
The complete multi center magnetic solutions  have been found recently
\cite{BeKallosh} in the form of  the  T-covariant magnetically charged
solutions of the heterotic string, defined by 28 harmonic functions. They have
one half of unbroken space-time supersymmetry of the heterotic string theory.

The interplay between the space-time supersymmetries and the world-sheet
supersymmetries for the BPS states was first studied in connection with the
heterotic instantons and solitons by Callan, Harvey and Strominger \cite{CHS}.
The analysis was performed for the five-branes and related to them H monopoles
\cite{Khuri}, \cite{GHL}. Now the large variety of more general magnetic
solutions is available for which  the interplay between the  space-time  and
the world-sheet supersymmetries was not studied yet.

The purpose of this paper is to study the generic class of the monopole
solutions of the heterotic string theory.   This means that we would like to
find the exact ten-dimensional supersymmetric solutions which  become 
monopoles
of the  four dimensional theory  upon dimensional reduction. Some of them are
expected to be massive, some massless.

One of the purpose of such uplifting of the four-dimensional monopoles was to
study the issue of anomaly related $\alpha'$ corrections to these 
monopoles and the corresponding world-sheet supersymmetric sigma models.
Another reason was to understand better the massless monopoles.
There was no information available about the behavior of the massless 
monopoles
under T duality. Moreover, the T self-dual solutions in this class is already
known \cite{BKO}: the uplifted
$a=1$ extreme massive magnetic black holes have such property. It seemed
unlikely that both $a=1$  as well as the massless black holes
can be both T self dual. Thus, we wanted to clarify what happens with massless
monopoles under T duality.

So far three types of monopole solutions of the heterotic string theory with
half of unbroken supersymmetry  were known to be exact.
For all of them the non-trivial part is a 4-dimensional
Euclidean manifold or the 10-dimensional manifold with   5 flat spatial
directions.

i) The first type, known in the literature as H-monopoles was first discovered
by Khuri \cite{Khuri}. The embedding of the spin connection into the gauge
group required the presence of the non-Abelian field in the solution.
At the time of their discovery  these solutions were interpreted as 
non-Abelian
monopoles. Soon after this work  Gauntlett, Harvey and Liu \cite{GHL} have
established the relation between these monopole solutions and the five-brane
solutions. They have also observed that the exact stringy monopoles of
\cite{Khuri} are actually not non-Abelian monopoles since the non-Abelian
vector fields fall down faster than the monopole field would. Rather they are
monopoles of a $U(1)$ group resulting from the compactification of the
antisymmetric tensor field  $B$, which  explains why these solutions 
are called H-monopoles.
The world-sheet supersymmetry of this solution, including the non-Abelian
fields, is known to be (4,4) which provides the  proof of  the absence of
$\alpha'$
corrections \cite{CHS}, \cite{HOWE}. The relation between H-monopoles and
extreme
$a= \sqrt 3$ magnetic black holes was realized in \cite{DuffKMR}.

ii) The second type of known monopoles \cite{Bian} if  treated in the same
spirit has the  right to be called F-monopoles.
Those solutions are  stringy
instantons with a constant dilaton, vanishing 3-form H and self-dual 
curvature
in the four-dimensional Euclidean subspace of the five-dimensional Minkowski
geometry.  They represent the Asymptotically Locally Euclidean (ALE)
gravitational
instantonic backgrounds coupled to gauge instantons
through the so--called  ``standard embedding''.
These solutions were found to be T-dual partners of the H-monopoles
\cite{Bian}. The non-Abelian fields are required to be present in the 
solution for the exactness.
The relevant non-Abelian fields also fall down as the dipole rather than a
monopole field, only the $U(1)$ field  $F=dA$ has a magnetic charge. 
The $U(1)$
field  $A$  originates  not in the antisymmetric tensor field but in the
non-diagonal component of the metric in the uplifted solution. The 
world-sheet
supersymmetry of this solution, including the non-Abelian fields, was found to
be  (4,4). These solutions from the point of view of the four-dimensional
geometry may be also associated with the extreme magnetic $a= \sqrt 3$ black
holes.

iii) The third type of exact magnetic solutions of the heterotic string is the
uplifted $a=1$ magnetic extreme black holes \cite{Gib}, \cite{GHS} ,
supplemented by the proper non-Abelian field for the exactness \cite{KO}. They
were called ``exact $SU(2)\times U(1)$ stringy black holes". Besides one
Abelian vector field $U(1)$ they had a non-Abelian $SU(2)$ vector field.
These solutions were found to be T self dual \cite{BKO}. In the spirit of
giving to monopoles the name according to the name of the gauge fields with
magnetic charges, this solution can be called F=H monopole. The world-sheet
supersymmetry of this solution, including the non-Abelian fields, was 
found to
be  (4,1) \cite{KO}, \cite{GK} which is sufficient to prove the absence of
$\alpha'$ corrections \cite{HOWE}.

\noindent In short, the first and the second type of monopoles related by T
duality are
\begin{equation}
(F_{\rm magn}=0, \; H_{\rm magn} ) \quad  \Longleftarrow T  \Longrightarrow
\quad    (F_{\rm magn}, \;  H_{\rm magn} =0)
\end{equation}
The third one is T self  dual
\begin{equation}
(F_{\rm magn}= H_{\rm magn}) \quad  \Longleftarrow T  
\Longrightarrow    \quad (H_{\rm magn}= F_{\rm magn})
\end{equation}

This picture of heterotic monopoles is obviously incomplete. One may
expect to find supersymmetric monopoles with 6 $U(1)$ fields H and
with 6 $U(1)$ fields F for the heterotic string compactified on a
6-dimensional torus. Those are solutions which we have found. Since
they have both F and H fields with the proper fall off at infinity,
and they interpolate between all three types of monopoles presented above,
we call them F \& H monopoles. Under the S duality our F \& H
monopoles transform into the electrically charged solutions.  The
electrical charge of the F-fields originates in the magnetic charge
of the H fields and vise versa. In particular, solutions with only
electric F fields become H monopoles and the electric solutions with
only H fields become F monopoles.
\begin{eqnarray}
(F_{\rm el}, \; H_{\rm el}=0) \quad  \Longleftarrow &S&  \Longrightarrow
\quad  (F _{\rm magn} =0, \; H_{\rm magn} )\nonumber\\
\nonumber\\
(F_{\rm el}=0, \; H_{\rm el}) \quad  \Longleftarrow &S&  \Longrightarrow
\quad  (F _{\rm magn} , \; H_{\rm magn}=0 )
\end{eqnarray}

Having these solutions with unbroken supersymmetry in the leading
approximation we may address the problem: which of these solutions are
exact? We will find a simple answer: all of them, with $SO(9)$ gauge
group for embedding of the spin connection. (For all of those with
$SO(8)$ gauge group the enhanced world-sheet supersymmetry takes
place). For all solutions which we will find, the non-Abelian part
always falls off faster than a monopole. The $SO(9)$ vector field far
away from the core falls off as $V ^{YM} \sim 1 / r^2$  and hence
the corresponding field strength as $\sim 1 / r^3$,  whereas the
Abelian field strength $F_{ij}$ and $H_{ij}$ fall off as $\sim \epsilon
_{ijk} x^k / r^3$.  Therefore the name F \& H monopoles remains valid
for these solutions even after they have been promoted to the exact
one. The world-sheet supersymmetry of F \& H monopoles will be found to
be at least (4,1) which is sufficient to prove the absence of $\alpha'$
corrections.

Many new features of the F \& H monopoles with none of F or H vanishing or
equal to each other, can be seen already at the level of one F and one 
H field,
i.e. at the level of the  solutions which is non-trivial only on the 
Euclidean
four manifold.  In particular we will study the massless monopoles upon
uplifting and the issues of T duality for this case and the structure of the
non-Abelian vector fields.

The paper is organized as follows. In Sec. 2 we present the most 
general known to us
solution of the heterotic string theory  in ten dimensions which is
supersymmetric and magnetically charged asymptotically flat solutions upon
dimensional reduction to four dimension. At this stage we consider only the
leading order string equations and do not study the  $\alpha'$ corrections;
there are no non-Abelian fields present.
However we have 6 $U(1)$ fields H and  6 $U(1)$ fields F as promised. 
In Sec. 3
we study the issue of the exactness of the general solution.  We  calculate
the spin connection for the uplifted monopole  solution and find how 
to promote
it to the exact one.  We find that the holonomy group of the spin connections
of monopole solutions is $SO(9)$. This comes as a nice surprise since the
electric partners of some of our monopoles have a holonomy group in the
non-compact part of the Lorentz group \cite{KO}, \cite{HT}.  Therefore the
issue of exactness of these electric solutions is not clear. However, all
magnetic solutions are fine and can be made exact by supplementing 
them by the
non-Abelian fields. In Sec. 4 we study the world-sheet supersymmetric sigma
models. For the most general $M^9$ monopoles we find $(1,1)$ 
supersymmetry. To
get the extended ones we study the $M^8$ monopoles and find (4,1) or (4,4)
supersymmetry. In Sec. 5 the $M^4$ monopoles are studied in detail. 
Finally in
the Appendix A we have the spin connections and the details about the 
holonomy
group of the monopoles. In Appendix B   we focus on subtleties of a
multi-monopole solutions with more than two centers.

\section{Heterotic monopoles}

The leading order heterotic string equations can be derived from the
following Lagrangian (we write the 10d fields with a hat):
\begin{equation}
S \sim \int dx^{10} \sqrt{\hat{G}} e^{-2\hat{\phi}}
[  \hat{R} + 4 (\partial \hat{\phi})^2 - \frac{1}{12}\hat{H}^2 ]
\label{10daction}\end{equation}
This action is the bosonic part of the pure $d=10, N=1$ supergravity.
We do not add any Abelian vector fields which would be responsible for
the 16 vector multiplets in toroidal compactification of the heterotic
string on ${\bf T}^6$. From this action we would get only 6 vector
multiplets in $d=4$. The reason for not working from the beginning
with Abelian vector multiplets in addition to the gravitational
multiplet is that we will need the non-Abelian vector multiplets in
$d=10$ to keep supersymmetry with account of quantum corrections.  Let
us start with the solution of this 10-dimensional theory and later we
will discuss the corresponding 4-dimensional theory.

\noindent
{\em A. Solution in $D=10$}

\noindent
 We assume that the fields depend only
on three coordinates ($x^i$) and denote the internal 6 coordinates by
$x^{\alpha}$. Thus we are looking for the static solutions with isometries in
all 6 internal directions. All fields are constructed out of 12 harmonic
functions.

The 10d metric is then given by\footnote{We use the notation
$\chi^L_{(\alpha} \chi^R_{\beta)}= \frac{1}{2} (\chi^L_{\alpha}
\chi^R_{\beta} +\chi^L_{\alpha} \chi^R_{\beta})$ and
$\chi^L_{[\alpha} \chi^R_{\beta]}= \frac{1}{2} (\chi^L_{\alpha}
\chi^R_{\beta} -\chi^L_{\alpha} \chi^R_{\beta})$.
}
\begin{equation}
\begin{array}{l}
\hat{ds}^2 = - dt^2 + e^{-4U} dx^i dx^i + (dx^{\alpha} + 
A^{(1)\alpha}_i dx^i)
 \,G_{\alpha\beta}\,(dx^{\beta} + A^{(1)\beta}_i dx^i) \\
\nonumber\\
e^{-4U} = 2 (|\vec{\chi}^R|^2 - |\vec{\chi}^L|^2) \quad , \quad
G_{\alpha\beta} = \delta_{\alpha\beta} - \frac{2 \chi^L_{(\alpha}
  \chi^R_{\beta)}}{|\vec{\chi}^R|^2 + (\vec{\chi}^R \vec{\chi}^L)}
\end{array}
\label{monopoles}\end{equation}
where $\vec{\chi}^{R}$ and $\vec{\chi}^{L}$ define  the
12-dimensional  harmonic  $O(6, 6)$-vector
\begin{equation}
\vec{\chi}(x) = \left( \begin{array}{c} \vec{\chi}^{L} (x)\\
\vec{\chi}^{R}(x)
             \end{array} \right)
                    \ , \qquad
 \partial_i  \partial_i  \vec{\chi}(x) =0 \ .
\label{magn}\end{equation}
For the dilaton we find
\begin{equation}
e^{-2 \hat{\phi}} = e^{2U} \, \frac{1}{\sqrt{\det G}} =
\sqrt{2} \, e^{4U} \, \frac{|\vec{\chi}^R|^2 + (\vec{\chi}^R 
\vec{\chi}^L)}{ |\vec{\chi}^R|}
\end{equation}
The 10d antisymmetric tensor components are given by \footnote{For the multi
monopole solution of the generic type with more than two centers there is a
subtlety concerning the status of $B_{\underline {ik}}$ terms. This will be
discussed in Appendix B.}.
\begin{equation}
\begin{array}{l}
\hat{B}_{\alpha\beta} = \frac{2 \chi^L_{[\alpha}
  \chi^R_{\beta]}}{|\vec{\chi}^R|^2 + (\vec{\chi}^R \vec{\chi}^L)}\\
\nonumber\\
\hat{B}_{\alpha\mu} = A^{(2)}_{\mu\alpha} + \hat{B}_{\alpha\beta}
A^{(1)\beta}_{\mu}
\end{array}\label{antisym}
\end{equation}
Both Kaluza-Klein gauge fields are pure magnetic and given by
\begin{equation}
\left(\begin{array}{c} F^{(1)}_{ij}\\ F^{(2)}_{ij} \end{array} \right) =
\sqrt{2} \epsilon_{ijm}\partial_m \left(\begin{array}{cc} 1 & 1\\
-1 & 1 \end{array} \right)\left(\begin{array}{c} \vec{\chi}^L \\
\vec{\chi}^R \end{array}\right)= \sqrt{2} \epsilon_{ijm}\partial_m
 \left(\begin{array}{c} \hskip 0.3 cm \vec{\chi}^L+  \vec{\chi}^R \\
-\vec{\chi}^L +\vec{\chi}^R \end{array}\right)
\label{FH}\end{equation}
The first six vector field strengths
\begin{equation}
F^{(1 ) \alpha}_{ij} =  \partial_i A^{(1 )  \alpha}_j - \partial_j A^{(1  )
\alpha}_i \equiv \vec F
\end{equation}
are build out of the non-diagonal component of the metric $g_{i}{}
^{\alpha}= A^{(1)\alpha}_i $.  When this magnetic field is present
in the solution we will call it F part of the monopole solution.
The second set of magnetic vector fields is based on the
antisymmetric tensor fields
\begin{equation}
F^{(2 ) }_{ij \,\alpha } =  \partial_i A^{(2 )  }_{j \, \alpha} -
\partial_j A^{(2  ) }_{i \, \alpha} \equiv \vec H \ .
\end{equation}
In the simplest case when $\hat{B}_{\alpha\beta}$ or  $A^{(1  ) \alpha}_i $
fields are absent
\begin{equation}
F^{(2 ) }_{ij  \,\alpha } =  \partial_i \hat {B}_{  \alpha \, j } - 
\partial_j \hat {B}_{\alpha \, i} \ ,
\end{equation}
as it follows from eq. (\ref{antisym}). Thus we will call  the non-vanishing
magnetic charges in the $F^{(2 ) }_{ij} $ sector  the H part of the solution.

All solutions above have one half of $d=10, \; N=1$ supersymmetry
unbroken.  This follows from the fact that we have found these
solutions by performing the supersymmetric uplifting of the BPS
solutions of $d=4, \; N=4$ theory \cite{BeKallosh}. For the choice of
asymptotically flat configurations which we adopt in this paper, the
supersymmetric uplifting can be performed by using the procedure and
notation of Maharana-Schwarz theory \cite{MS} and developed by Sen
\cite{Sen}. If we would be interested in configurations which are not
asymptotically flat, one would have to switch to the more general case
of dimensional reduction and use the work by Chamseddine \cite{Cham}.
This would give the correct assignment of the fields to various
supersymmetric multiplets.

As was already explained, the reason we call solutions 
(\ref{monopoles}) F\& H
monopoles is the fact that they become magnetically charged solution with two
sets of vector fields, F and H, upon dimensional reduction. Specifically,
  the $O(6,6)$-invariant bosonic action  in the form of
Maharana-Schwarz \cite{MS}
and  Sen \cite{Sen} is
\begin{eqnarray}\label{action}
S &=& {1\over16\pi} \int d^4 x \sqrt{-\det G} \, e^{-2 \phi} \, \Big[ R_G + 4
G^{\mu\nu}
\partial_\mu \phi \partial _\nu\phi +{1\over 8} G^{\mu\nu} Tr(\partial _\mu
{\cal M} L\partial_\nu  {\cal M} L)
\nonumber \\
&&  -{1\over 12}  (H_{\mu\nu\rho})^2
 -  \frac{1}{4} G^{\mu\mu'} G^{\nu\nu'} F^a_{\mu\nu} \, (L {\cal M} L)_{ab}
\, F^b_{\mu'\nu'} \Big] \, .
\end{eqnarray}
This is a bosonic part of $N=4$ supergravity interacting with six 
$N=4$ vector multiplets. There is one
vector field in each vector supermultiplet and 6 vector fields in 
supergravity multiplet.
 We are looking for the solutions in which all vector fields are magnetic and
there are no axions $H_{\mu\nu\rho}=0$. The magnetic potentials for 6
graviphotons are $ \chi^R_{\alpha}$ and the magnetic potentials
for the six vector multiplets are
$\chi^L_{\alpha}$. They are all harmonic
functions, as shown in eq. (\ref{magn}).

\noindent
{\em B. Solution in $D=4$} \nopagebreak

\noindent
The  four-dimensional  supersymmetric solution corresponding to the uplifted
supersymmetric solution in eq. (\ref{monopoles}) is given by \cite{BeKallosh}
\begin{equation} \label{monopole4}
\begin{array}{ccc}
ds^2_{\rm str}= -  dt^2 + e^{-4 U} d\vec{x}^2\ , &&  e^{-4U} = 2 \,\chi^T L
\chi
                         = e^{4\phi} =2 (|\vec{\chi}^R|^2
- |\vec{\chi}^L|^2)\ , \\
\nonumber\\
\nonumber \\
{\cal M}= {\bf 1}_{12} + 4 e^{4U} \left( \begin{array} {cc}  \chi^L_{\alpha}
\chi^L_{\beta}&  \chi^L_{\alpha} \chi^R_{\beta}\\
\nonumber\\
  \chi^R_{\alpha} \chi^L_{\beta} &  \xi \chi^R_{\alpha} \chi^R_{\beta}
\end{array} \right) , & &\left(\begin{array}{c} F^{(L)}_{ij}\\ F^{(R)}_{ij}
\end{array} \right) =
2  \epsilon_{ijm}\partial_m  \left(\begin{array}{c} \vec{\chi}^L \\
\vec{\chi}^R \end{array}\right)\end{array}
\end{equation}
where $\xi={ |\vec{\chi}^L|^2\over |\vec{\chi}^R|^2}$.
The canonical four-dimensional metric is
\begin{equation}
ds^2_{\rm can}=  -e^{2 U}dt^2 + e^{-2 U} d\vec{x}^2 \ .
\end{equation}
The one center solutions can be taken in the simplest form
\begin{equation}
\vec \chi^L = \frac{ \vec P_{\rm vec}}{2 |\vec x|} \ , \qquad \vec \chi^R  =
{\vec n \over \sqrt 2} + \frac{  \vec P_{\rm
gr}}{2 |\vec x|} \ , \qquad \vec n^2 =1 \ , \qquad \vec n \cdot \vec P_{\rm
gr} \geq 0
\end{equation}
For this spherically symmetric solution  $ |\vec x| = \sqrt {(x^1)^2 + 
(x^2)^2
+ (x^3)^2} = r $. It may be useful to rewrite the expressions for
$\vec F$ and $\vec H$  in an explicit form in terms of magnetic charges.
\begin{equation}
\left (\matrix{
\vec F_{ij}\cr
\cr
\vec H_{ij}\cr
}\right ) ={1\over  \sqrt 2}  \epsilon_{ijm}{x^m \over |\vec x|^3}  \left
(\matrix{
\vec P_{\rm
gr} +\vec P_{\rm
vec } \cr
\cr
\vec P_{\rm
gr } - \vec P_{\rm
vec }\cr
}\right ) =  \epsilon_{ijm}{x^m \over |\vec x|^3}  \left (\matrix{
\vec P_{F} \cr
\cr
\vec P_{H
 } \cr
}\right )
\end{equation}
where we have defined
\begin{eqnarray}
\vec P_{F}&\equiv & {1\over  \sqrt 2} (\vec P_{\rm
gr} +\vec P_{\rm
vec })\\
\vec P_{H}&\equiv &{1\over  \sqrt 2}   ( \vec P_{\rm
gr} -\vec P_{\rm
vec })\ .
\label{FHcharges}\end{eqnarray}

Using  S duality \cite{Sen}, \cite{BeKallosh} one can convert the monopole
solutions into electrically charged
ones.  The  electric  solution is given by the following formula
\begin{equation}   \label{electr}
ds^2_{\rm str}= -  e^{ 4 U}dt^2 +  d\vec{x}^2 \ , \qquad  e^{4 U}=e^{4 
\phi} \ ,
 \qquad E_i ^{(a)}  = {1\over  2}  e^{4 U} ({\cal M} L)_{ab}\, \partial_i
\chi^b\ ,
\end{equation}
where $U$ and ${\cal M}$ are defined in eq. (\ref{monopole4}) and $E_i 
^{(a)}$ is the  electric
field.
$E_i ^{(a)} = F_{ti}^{(a)}$. One can see from this formula that the reason why
the vector multiplet charge changes the sign during S duality is the 
following. The asymptotic value of the matrix  ${\cal M} L$ is
\begin{equation}
{\cal M} L= \pmatrix{
-1 & 0 \cr
0 & 1 \cr
} \ .
\end{equation}
Therefore the upper part related to vector multiplets $\chi^L$ changes 
the sign
whereas the lower part related to $\chi^R$ does not change the sign. 
It follows that
the magnetic charges of the graviphotons become the electric charges of the
graviphotons
\begin{equation}
\vec P_{\rm
gr} \; \Longleftarrow S \Longrightarrow \;  \vec Q_{\rm
gr}
\end{equation}
and the magnetic charges of the vector multiplets become  electric charges of
the vector multiplets with the opposite sign
\begin{equation}
\vec P_{\rm
vec }  \; \Longleftarrow S \Longrightarrow  \; - \vec Q_{\rm
vec}
\end{equation}
Thus S duality trades  F  for H fields  and vise versa.
\begin{eqnarray}
\vec F_{\rm magn}   \; \Longleftarrow &S& \Longrightarrow \;  \vec 
H_{\rm el}\\ \nonumber\\
\vec H_{\rm magn}
  \; \Longleftarrow &S& \Longrightarrow \; \vec F_{\rm el}
 \end{eqnarray}

We have presented in \cite{BeKallosh} the classification of the monopoles via
the classification
of their S dual electric partners for which the relation to the elementary
string excitation is available \cite{Sen}, \cite{DuffRum}.
Since all solutions which we consider are supersymmetric  the right-moving
oscillator modes have
$N_R = \frac{1}{2}$. For the left-moving part we obtain
\begin{equation}
N_L - 1 = {1 \over 2} (Q_{\rm gr}^2 - Q_{\rm vec}^2)  \qquad M^2 = {1\over 2}
Q_{\rm gr}^2\ .
\end{equation}
 The electric solution (\ref{electr}) describes the following states:

\noindent 1) $N_L=0$  massive  and massless white holes
 \newline
2) $N_L=1$  extremal $a=\sqrt{3}$ black holes.
\newline
3) $N_L \geq 2$  discrete set of extremal black holes
   (for $  M^2 =$ $N_L -1$ they reduce to  $a=1$ black holes).

Our magnetic configurations (\ref{monopoles}) can be also associated with
various values of $N_L$ via the relation (we consider one center 
solution here) \begin{equation}
N_L - 1 = {1 \over 2} (Q_{\rm gr}^2 - Q_{\rm vec}^2) = {1\over 2}  (P_{\rm
gr}^2 - P_{\rm vec}^2)=  (\vec P_{F} \cdot \vec P_{H}) \ ,    \qquad M^2 =
{1\over 2} P_{\rm gr}^2\ .
\end{equation}

\noindent 1) $N_L=0$   massive  and massless monopoles

\begin{equation}
\vec P_{F} \cdot \vec P_{H} = -1 \ ,  \hskip 1 cm  (P_{\rm
gr}^2 - P_{\rm vec}^2) =-2 \ ,
\end{equation}
where the massless limit is given by
\begin{equation}
\vec P_{F} + \vec P_{H} = 0\ ,        \qquad   \vec P_{\rm
gr} =0 \ ,    \qquad P_{\rm vec}^2 = 2 \ .
\end{equation}
 \newline
2) $N_L=1$   monopoles \begin{equation}
\vec P_{F} \cdot \vec P_{H} = 0\ , \hskip 1 cm  P_{\rm
gr}^2  = P_{\rm vec}^2 \ .
\end{equation}
Obviously H monopoles satisfy this constraint since for them $\vec F =0$.
However, any solution with $\vec F \cdot \vec H = 0$ also represents  an
$N_L=1$ state. In particular, $\vec H =0$ or any solution with non-vanishing
but orthogonal  $\vec F$ and $ \vec H $ also belongs to  $N_L=1$ state.

\noindent 3) $N_L \geq 2$  monopoles
\begin{equation}
\vec P_{F} \cdot \vec P_{H} \geq  1 \ .
\end{equation}
In the special case
 $M^2 = N_L -1$ they reduce to  $a=1$ extreme magnetic black holes with
\begin{equation}
\vec P_{F} - \vec P_{ H} = 0 \ , \hskip 1 cm  \qquad  \vec P_{\rm
vec } =0 \ .
\end{equation}
The remarkable feature of all F \& H monopoles with $N_L \geq 2$ was observed
in \cite{BeKallosh}.
 In the stringy frame in the four-dimensional geometry they
are completely non-singular solutions. We get for the metric
at  $r \rightarrow 0$
\begin{equation}
ds^2_{str} = -dt^2 + 2(|\chi^R|^2 - |\chi^L|^2) d\vec{x}^2 \rightarrow
-dt^2 + d\rho^2 +  (\vec P_{F} \cdot \vec P_{H}) d^2 \Omega
\end{equation}
where $d\rho = \sqrt{\vec P_{F} \cdot \vec P_{H}} dr/r$.
Hence, in this limit the 4-dimensional solution is given by a
bottomless throat
($M_2 \times S_2$), with the radius squared given by the scalar product of
both charge vectors. By identification with the string excitation
we find that the radius squared   has to be quantized
($\vec P_{F} \cdot \vec P_{H} =( N_L -1 )\geq  1$) and we get the limiting
metric in the form
\begin{equation}
ds^2_{str} \rightarrow
-dt^2 + d\rho^2 +  (N_L-1) d^2 \Omega
\end{equation}
Obviously, for any $N_L=1$ ($\vec P_{F} \cdot \vec P_{H} =0)$
excitation the radius of the throat vanishes (singularity) and for $N_L=0$
the throat shrinks to zero already at finite $r=r_c > 0$.
The expression for the scalar curvature
was calculated in \cite{BeKallosh}. Using our new notation which
focus on the presence of the F {\em and} H charges, defined in
eqs. (\ref{FHcharges}) we have
\begin{equation}
R_{\rm str} =   {2(4 \vec P_{F} \cdot \vec P_{H})^2 - 8 M (4 \vec P_{F} \cdot
\vec P_{H}) r + 4(6 M^2  -  (4 \vec P_{F} \cdot \vec P_{H}))r^2
\over  \left ((4 \vec P_{F} \cdot \vec P_{H})  +4Mr + r^2\right )^3}\ .
\label{curvature}\end{equation}
Looking on the denominator of this expression one can see that the
solution is non-singular for all positive
$(\vec P_{F} \cdot \vec P_{H})$. In terms of dual string states
this scalar product has to be a positive integer defined by
$(\vec P_{F} \cdot \vec P_{H}) =  (N_L-1)$ for
$N_L \geq 2$. The maximum curvature for these solutions is reached at $r=0$
and is equal to
\begin{equation}
R_{\rm str}(0) =   {1\over  2  (N_L-1)}\ .
\label{maxcurvature}\end{equation}

\section{Stringy $\alpha'$ corrections to F \& H monopoles}

There is a strong belief that supersymmetry established at the 
classical level
will be preserved with the account of the quantum corrections in absence of
anomalies.  However, when anomalies are present the preservation of
supersymmetry and the issue of BPS states in general are not clear.  In some
particular situations one can study the quantum corrections to the
supersymmetry transformations which are due to anomalies. This has 
been  worked
out specifically for the anomaly related $\alpha'$ corrections in the 
heterotic string theory.
The relevant supplement to the 10-dimensional action  (\ref{10daction})
includes the Yang-Mills field and the  generalized curvature coupling
(with torsion)
\begin{equation}
\alpha' \left ( Tr  (R_{-})^2 - Tr  F^2 \right )
\label{alpha}\end{equation}
where the Trace operation in the $Tr  (R_{-})^2$ is over the non-compact
Lorentz group $SO(1,9)$ and the one in the Yang-Mills $Tr F^2$ is over the
compact $SO(32)$ or $E_8\times E_8$. The BPS configuration which solves
the leading order equations supplies the information about the generalized
curvature, which gives  the first term  in eq. (\ref{alpha}).  When the
generalized ten-dimensional curvature $R_{- AB}{}^{CD}$ has a nonvanishing
value in the non-compact direction of the Lorentz group, i.e.
\begin{equation}
R_{- AB}{}^{0D} \neq 0
\end{equation}
there is no possibility to use the standard procedure of the spin embedding
into the gauge group. However, when
\begin{equation}
R_{- AB}{}^{0D} =0 \ ,
\end{equation}
the spin connection can have  a maximum  holonomy group $SO(9)$ which is
compact. In such situation one can use the standard procedure of embedding of
the spin connection into the gauge group. The advantages of this are

1. The anomaly related $\alpha'$ corrections to the equations of motion in the
effective theory, to the action and to the {\it  space-time supersymmetry}
rules vanish. The detailed description of the procedure can be found in
\cite{BKO1}.

2. In terms of the world-sheet supersymmetry embedding of the spin connection
into the gauge group means the following: one can start with the (1,0)
supersymmetric model and enhance this supersymmetry to (1,1) {\it world-sheet
supersymmetry}. This is believed to be necessary for avoiding chiral 
anomalies
in the left-right asymmetric case. The details of this procedure can be found
in \cite{CHS}.

The procedure of correcting the supersymmetric solutions via spin embedding
into the gauge group was applied before to many BPS solutions, starting with
the symmetric version of the five-brane  \cite{CHS}. The same procedure was
also applied to H monopoles \cite{Khuri},  to F monopoles \cite{Bian} and to T
self dual monopoles \cite{KO}. Typically for
all these solutions the Yang-Mills field  to be added to the solutions 
was part of the $SO(4)$ gauge theory.
In application to the gravitational waves \cite{BKO1}  and to the generalized
fundamental strings  \cite{BEK} the corresponding gauge group was found to be
$SO(8)$.

It was known however, that  the uplifted electrically charged $a=1$ 
black holes
have the non-compact holonomy group of the generalized spin connections
\cite{KO}.
In more general models, chiral null models of Horowitz and Tseytlin \cite{HT},
the holonomy group of the generalized spin connections is also not compact
\cite{HT} since it includes the non-compact Abelian subgroup of the Lorentz
group. For these solutions the possibilities of   restoring the unbroken
supersymmetry of the classical solution in presence of $\alpha'$ corrections
are not clear.

For the generic F \& H monopoles the non-Abelian group was not known before,
since to find the gauge field one has to calculate the spin connection with
account of the torsion. We will first study the general solution and find that
the holonomy group of the generalized spin connection is  the compact group
$SO(9)$.  This solution has a non-trivial metric in all 9 directions but time,
therefore  we will sometimes call it ${\bf  M}^9$.
We will describe this most general solution and the corresponding (1,1)
supersymmetric sigma model.
Afterwards we will focus on the slightly less general solution which can be
embedded into $SO(8)$ and whose geometry is non-trivial on ${\bf  M}^8$
Euclidean manifold.
 For this solution we will study the extended supersymmetries on the
world-sheet.  The reason for this is that to enhance the single world-sheet
supersymmetry (1,0) to (2,0)  one has to consider a manifold of even dimension
and to enhance it to (4,0) one need a manifold of dimension $4n$ with the
integer $n$.

Finally we will
perform the detailed study of F \& H monopoles with account of $\alpha'$
corrections in the simplest case of  ${\bf  M}^4$ monopoles which have a
non-trivial geometry in the four-dimensional Euclidean space. The relevant
gauge group will be again the $SO(4)$.

In this section we will have to introduce the set of notations suitable for
dealing with vielbeins and spin connections. We will use the base manifold
coordinates on ${\bf M}^{10}$  and denote them $x^M = \{ t ; 
x^{\underline i} =
( x^1, x^2, x^3) ;  x^\alpha =(x^4, \dots , x^9) \}$. The tangent 
space will be introduced via the zehnbeins
\begin{equation}
\hat E^A = \hat E^A{}_M dx^M \ , \qquad \hat E^A = \{ e^0; e^i; E^a\} \qquad
i=1,2,3;  \; a= 4, \dots , 9
\end{equation}
The uplifted monopoles metric (\ref{monopoles}) can be rewritten in the form:
\begin{equation}
ds^2 = - (e^0)^2 + (e^i)^2 +  (E^a)^2  = \hat E^A \eta _{AB} \hat E^B \,
\end{equation}
where
\begin{eqnarray}
e^0&=& dt \label{timelike}\\
e^i &=& e^{-2U} dx^{\underline i} \delta _{\underline i} {}^i\\
E^a &=& E^a{}_\alpha (dx^\alpha +  A ^{(1){\alpha}}{} _{\underline i }d
x^{\underline i})
\end{eqnarray}
and the tangent space metric is $\eta_{AB} =\{ -1, +1, \dots , +1\}$.
The explicit expressions for the six-dimensional vielbein $E^{a}{}_{\alpha}$
and its inverse $E^{\alpha }{}_{a}$ is defined in terms of our 12 harmonic
functions and can be found in the Appendix A. Thus it is clear from eq.
(\ref{timelike}) that the most general monopole solution in this class has a
non-trivial nine-dimensional Euclidean manifold
${\bf  M}^9$.
The spin connection one-form will be defined as
\begin{equation}
W_{AB} \equiv W_{AB, C} \hat E^C
\end{equation}
with the standard definition of $W_{AB, C}$ in terms of zehnbeins and its
derivatives.
To build the generalized spin connections we need also the tangent 
space 3-form $H_{ABC}$.

The objects of interest are the torsionful spin connections
\begin{equation}
\Omega_{ \pm AB}= (W_{AB, C}  \pm  H_{ABC} )E^C \equiv W_{AB} \pm H_{AB}
\end{equation}
Our tangent space group is the Lorentz group $SO(1,9)$ with 45 
generators, 9 of
them are boosts $B^I$ and 36 are $SO(9)$ rotations $M^{IJ}$:
\begin{equation}
M^{AB} = \left \{ B^I \equiv M^{[0I]} , \; M^{IJ}  \right \} \ , \qquad I,J =
1,\dots ,9.
\end{equation}
The boosts generators $B^I   = \{M^{[0i]}, \; M^{[0 a]}  \} $ are responsible
for the non-compact   directions whereas  $M^{IJ} = \{M^{[ij]}, \; 
M^{[ab]}, \; M^{[ia]} \}$ are  responsible for the $SO(9)$ rotations.
In principle the generalized spin connections may take values in any part of
the Lorentz group, in compact part as well as in a non-compact one.  We have
found that for all F \& H monopoles  both the metric spin connections $W_{AB} $
as well as torsion part of the spin connection  $H_{AB}\equiv H_{ABC} E^C $
take values only in the $SO(9)$ rotation part of the Lorentz group.
\begin{equation}
\Omega_{ \pm 0I} =0 \qquad  \Omega_{ \pm IJ} \neq 0 \ .
\end{equation}
The holonomy algebra of the generalized spin connections is the algebra
generated by the $M^{IJ}$ and the holonomy group is $SO(9)$. We have 
performed
the explicit calculation of the metric part of the spin connections 
adapting to
our case the well known formulas of Scherk and Schwarz \cite{SS}. The $SO(9)$
Yang-Mills  one-form field is given by the non-vanishing components of
$\Omega_{ - IJ}$ spin connection
$$V_{IJ} = V_{IJ, K}  E^K  =\Omega_{ - IJ} $$
The details of the calculation and the expression for $ \Omega_{ \pm  IJ}$
can be found in the Appendix A.
The net result for the Yang-Mills vector field is
$$V_{  IJ, \,0} =0 \qquad  V_{  IJ, \, K}  = (W_{IJ, \, K}  +  H_{IJK}) $$
The first two indices are the indices of the $SO(9)$ gauge group and 
the third one is a space-time index (in the tangent frame).
The zero space-time component of the tangent space vector field vanishes for
all solutions: this means that the non-Abelian field is also of a magnetic
nature.  It is therefore tempting  to find out if the $SO(9)$ field  
$V_{  IJ ,
K}$ carries any magnetic charge. For this purpose we note that far from the
core of the monopole (we study the one-center solution here) the  vielbeins
behave as
$$E^A{}_M \sim   c+ {c_1 \over r} + {c_2 \over r^2 } + \dots$$
Therefore the metric spin connections, which depends on  vielbeins and
derivatives of the vielbeins,  behave as
$$W_{IJ,\, K} \sim    {d_1 \over r^2} + {d_2 \over r^3 } + \dots$$
The same large distance behavior can be observed for the torsion part of the
spin connection. Indeed the curved space 3-form consists of the 
derivative of a 2-form field which behave as
$$B_{MN} \sim   f + {f_1 \over r} + {f_2 \over r^2 } + \dots$$
The 3-form therefore behaves as
$$H_{MNL} \sim   {e_1 \over r^2} + {e_2 \over r^3 } + \dots$$
Thus all F \& H monopoles have the non-Abelian vector field which comes from
the embedding of the spin connection into the gauge group $SO(9)$ and 
at large distances  falls off as
$$V_{IJ, K} \sim {a_2 \over r^2}+  {a_3 \over r^3 } + \dots$$
The Yang-Mills field strength would fall off as ${1\over r^3}$. The situation
is exactly the same as described in \cite{GHL} for the H monopoles: 
there is no
magnetic charge associated with the non-Abelian part of the solution, we have
only F \& H  magnetic charges associated with the Abelian vector fields, which
we have discussed above.

Since we have established that all  F \& H monopoles can be 
supplemented by the
non-Abelian $SO(9)$ field  via spin embedding one can address the issue of the
world-sheet supersymmetry of the heterotic string theory in the generic
monopole background.

  \section{World-sheet actions  for ${\bf  M}^9$ and ${\bf  M}^8$ string
monopoles}
After the four dimensional monopole solutions have been interpreted as
 solutions with unbroken supersymmetries of
the effective
action of the heterotic  string theory in critical dimension, one can 
construct a supersymmetric
sigma model in an uplifted monopole target space. The details for the special
case of the uplifted $a=1$ black holes can be found in \cite{GK}. The most
general monopole solutions with $SO(9)$ non-Abelian gauge field defined by 12
magnetic charges suggest the following $(1,1)$ supersymmetric sigma model
\cite{HOWE}.
\begin{equation}
I_{(1,1)} = \int d^2 z d^2 \theta  (G_{\underline{IJ} } + B_{\underline{IJ}})
D_+
X^{\underline{I}} D_-
X^{\underline{J}}\  ,
\label{superaction}\end{equation}
where the unconstrained $(1,1)$ superfield is given by
\begin{equation}
X^{\underline{I}} (x^{\underline{I}}, \theta^+, \theta^-) = x^{\underline{I}}
(z) + \theta^+
\lambda_+{}^I (z)
E_I{} ^{\underline{I}} (x)
 - \theta^-
 \lambda_-{}^I(z)   E_I{}^{\underline{I}} (x)   - \theta^+\theta^-
F^{\underline{I}} (z)\ .
\label{superfield}\end{equation}
This theory is defined in the Euclidean 9 dimensional manifold  ${\bf  M}^9$
given by the components of the $9\times 9$ sector of the monopole solution
(\ref{monopoles}),
(\ref{antisym}).

To find out the class of monopole solutions for which the extended
supersymmetry can be established we will limit ourself to the case
when one of the directions in the internal 6 manifold is flat. Let
$\chi_{9}{}^R = \chi_{9}{}^L = 0$. This solution is the one defined
in eq. (\ref{monopoles}) but characterized by ten harmonic functions
instead of twelve.  This monopole solution has very special
properties. The non-trivial background is  in ${\bf M}^8$ only
\begin{eqnarray}
G_{MN} dx^M dx^N &=& - dt^2 + (dx^9)^2 + e^{-4U} dx^i dx^i \nonumber\\
&+& \sum_4^8  (dx^{\alpha} +  A^{(1)\alpha}_i dx^i)
 \,G_{\alpha\beta}\,(dx^{\beta} + A^{(1)\beta}_i dx^i)
\label{monopole8}\end{eqnarray}
The 10d antisymmetric tensor components $B_{MN} =( B_{\alpha \beta} \ ,
B_{\underline i  \alpha}) \ , \alpha =4,\dots 8$
are given in eq. (\ref{antisym}).

We may rewrite the geometry  (\ref{monopole8}) of the ${\bf M}^8$ 
monopoles as the function of ten harmonic functions
\begin{equation}
 ds^2 = du dv  + G_{\underline {IJ}}(\chi_{\alpha}{}^L,  \chi_{\alpha}{}^R)
dx^{\underline I} dx^{\underline J}  \qquad {\underline I},  I = 1,\dots , 8
\qquad u=- t + x^9, \; v=t+x^9
\end{equation}
$$B_{\underline {IJ}} = B_{\underline {IJ}}(\chi_{\alpha}{}^L,
\chi_{\alpha}{}^R) \qquad  \alpha =4,\dots 8$$
and we may also see that only $B_{\underline {IJ}}$ are non-vanishing and are
given in eq. (\ref{antisym})
The main property of ${\bf M}^8$ monopole background is: one can prove that
the Green-Schwarz and Ramond-Neveu-Schwarz formulation of the superstring are
equivalent in this background. The light-cone action of one theory can be
transformed into the light-cone action of the other theory by converting
$SO(8)$ spinors into $SO(8)$ vectors.

To prove this we will study the conditions of such equivalence, as given by
Hull in \cite{Hull}.

i) The non-trivial part of the background in the stringy frame has to 
describe an 8 dimensional Euclidean manifold

ii) The background has to admit a sufficient number of supercovariantly
constant Killing spinors. This allows to construct at least 3  almost complex
structures in the right and/or in the left-moving sectors of the theory in
which Killing spinors exist.

The equivalence theorem proved by Hull \cite{Hull} and also the study of
supersymmetric sigma models by Hull and Witten \cite{HW}  and Howe and
Papadopolous \cite{HOWE} indicate also that  for some backgrounds for 
which in addition

iii) the Nijenhuis tensor vanishes, one may find the enhancement of
supersymmetry from 1 up to 4 in each of the right or left moving
sectors of the theory where these complex structures have been
found.  \vskip 0.6 cm All three conditions are met for ${\bf M}^8$
monopoles.  We will find that the world-sheet action in the ${\bf
  M}^8$ monopole background has an extended $(4,4)$ supersymmetry
for all solutions with $N_L=1$ and $(4,1)$ for the rest.  We may
again use the $(1,1)$ action as given in eq. (\ref{superaction}),
however now ${\underline I}, I = 1,\dots , 8$. Upon integration over
the fermionic variables and elimination of the auxiliary fields we
get
\begin{eqnarray}
S &=& \int d\tau d\sigma [ (G_{\underline {IJ}} + B_{\underline {IJ}})
\partial_z x^{\underline {I} }\partial_{\bar z}
x^{\underline {J}}+i \lambda_+{}^I (\nabla_{z}{}^{(+)} \lambda_+) ^I - i
\lambda_-{}^I
(\nabla_{\bar z}{}^{(-)} \lambda_-) ^I \nonumber\\
\nonumber\\
&-&{1\over 4} R^{(+)}_{IJ,KL} \lambda_+{}^I  \lambda_+{}^J
\lambda_-{}^K
\lambda_-{}^L + {1\over 4} R^{(-)}_{IJ,KL} \lambda_-{}^I  \lambda_-{}^J
\lambda_+{}^K
\lambda_+{}^L] \ .
\label{lagr2}\end{eqnarray}
The right(left)-handed fermions $\lambda_+{}^I$ ($\lambda_-{}^I$)
have covariant derivatives
with respect to torsionful spin connections $\Omega_{+}$ ($\Omega_{-}$).
The torsionful curvatures $R_{\pm} = d\Omega_{\pm} + \Omega_{\pm}\wedge
\Omega_{\pm}$ have the exchange properties
\begin{equation}
R^{(+)}_{IJ,KL}= R^{(-)}_{KL,IJ} \ .
\label{exch}\end{equation}
 For our monopoles with the non-Abelian $SO(8)$ fields  the torsionful
curvatures are given  by
the Yang-Mills field strength which due to spin embedding is equal to
$R^{(-)}_{IJ,KL}$
and  by the gravitational torsionful curvature $R^{(+)}_{IJ,KL}$,
which are related to each other by eq.\ (\ref{exch}).

This action has more of the extended supersymmetries for our monopole
solutions. Indeed, one can find the set of three almost complex 
structures  by
building them in terms of the bilinear combinations of  Killing 
spinors of our
background.  The corresponding commuting normalized   spinors are 
$\alpha^{\dot
p}, \beta^{(m)  \dot p},  m \leq 7$ where  $\dot p$ is the spinorial index of
the $SO(8)$. The complex structure $J_{KL}$
with all required properties was found in \cite{Hull} to be
\begin{equation}
J_{KL}{}^{(m)} =  \beta^{(m)}_{ \dot p} \sigma ^{\dot p \dot q}_{KL}
\alpha_{\dot q} \ .
\label{J}\end{equation}
Although the number of such almost complex structures can be as large 
as 7, we
are interested only to find at least two, the third one being defined by the
two. If  there are more than three, the target manifold is reducible. The
counting of complex structure proceeds as follows. For solutions with $N_L=1$
for which the square of the right-handed magnetic charge equals the square of
the left-handed magnetic charge, the theory can be embedded into the type II
superstring theory (or into $N=8$ supergravity at the level of the effective
four-dimensional action). This solution has left-right symmetry and therefore
it has one half of unbroken supersymmetry in both
left- and right-handed spinors, i.e. for our $\vec P_F \cdot \vec P_H=0$
monopoles we have the double set of almost complex structures. Thus for
$N_L=1$ monopoles one can expect  the enhancement of world-sheet supersymmetry
from the manifest one in eq. (\ref{superaction}) which is $(1,1)$ up 
to $(4,4)$
if all necessary properties of the complex structures will be established.

For all infinite tower of other solutions with $N_L =0, N_L =2, 3,
\dots, n$ the left-right symmetry is broken since  $ P_{R }^2 =
P_{L}^2 +2 (N_L - 1))$  and therefore $P_{R }^2 \neq P_{L}^2 $.
The space-time Killing spinors exist only in the right moving
sector, the left moving one does not have unbroken supersymmetries.
These solutions have one half of unbroken supersymmetry of the
heterotic string (and only one quarter of type II string).
Therefore one can construct the complex structures out of space-time
Killing spinors according to Hull's prescription (\ref{J}) only for
the right-moving modes. Therefore, for these solutions the expected
world-sheet  supersymmetry enhancement  is going to be $(4,1)$
under condition that the algebra of these extended supersymmetries
closes.

Now that we have established that  we do have enough of almost complex
structures,
since we deal with  configurations with unbroken space-time supersymmetries,
the crucial question remains whether the commutator of two supersymmetries
closes  for some of our monopoles or for all of them. This does not seem to be
a property of an
arbitrary background
even with unbroken supersymmetries. The right hand side of the commutator of
the first supersymmetry with the
one induced by the existence of a covariantly constant almost complex 
structure
 $J$ depends of the Nijenhuis tensor
\begin{equation}
N^K_{IJ} = J^L{}_I J^K {}_ {[J, L]}- J^L{}_J J^K {}_ {[I, L]} \ .
\end{equation}
In this expression comma means a derivative.
For a generic solution with unbroken supersymmetry the complex 
structure $ J^K {}_ J$
is covariantly constant, but not necessarily constant, which would force terms
like
$J^K {}_ {[J, L]}$  to vanish. Therefore  the Nijenhuis tensor  does not
vanish in general  and therefore one does not find the
enhancement of supersymmetry for any supersymmetric background. However, we
have found that for our ${\bf M}^8$ monopole solution the
Nijenhuis tensor  does vanish, which provide the closure of the algebra of the
extended supersymmetries on the world-sheet. The reason is that {\it all
magnetic solutions of the heterotic string have Killing spinors in the 
stringy frame which are constant}. This property of monopoles provides 
constant complex structures $ J^K {}_ J$
with
$J^K {}_ {[J, L]} =0$. Indeed,
in canonical frame
\begin{equation}
\epsilon^{\rm can}(x)  = e^{U(x) \over 2} \epsilon _{0} \ ,
\label{killing}\end{equation}
where $\epsilon _{0}$ is a constant spinor.
For all magnetic solutions with $U+\phi =0$ this means that the covariantly
constant spinor in stringy frame is a constant spinor!
\begin{equation}
\epsilon^{\rm str} = \epsilon _{0}
\end{equation}
This property was observed for the five-branes and H monopoles before
\cite{CHS}, \cite{Khuri}, where it was also used for the enhancement
of world-sheet supersymmetry.
For $a=1$ magnetic black holes it was also found \cite{US} that in the 
stringy
frame the Killing spinors exists globally since they are constant. Now 
we have
verified that for the total family of pure magnetic solutions in the stringy
frame the Killing spinors are constant. This can be done for example by
observing that the Killing spinor for the
$O(6,22)$  covariant electrically charged solutions was found by Peet
\cite{Peet} to be of the form (\ref{killing}). By S duality it follows 
that for
all magnetic solutions Killing spinors are constant in stringy frame.

Thus ${\bf M}^8$ monopoles (\ref{monopole8}) with $SO(8)$ non-Abelian gauge
fields
formulated in the stringy frame have  the following properties:

i) non-trivial 8 dimensional Euclidean geometry and unbroken space-time
supersymmetry  which exists globally: the Killing spinors as well as the
complex structures are constant

ii) $N_L=1$ solutions correspond to sigma models with $(4,4)$ world-sheet
supersymmetry; the GS formulation of the type II superstring theory with
manifest space-time supersymmetry is equivalent to the NSR form with the
world-sheet supersymmetry: the world-sheet  (1,1) spinor  $\lambda_{+}^I,
\lambda_{-}^I, $ which is also an $SO(8)$ vector is converted into the
space-time $SO(8)$ spinors $S^q,  S^{\dot q}$ using the normalized commuting
Killing spinors  $\alpha^{\dot p}, \alpha^{ p}$ of the monopole background:

\begin{equation}
\lambda_{+}^I = \alpha^{\dot p}  \gamma^I_{\dot p q} S^q \ , \qquad
\lambda_{-}^I = \alpha^{ p}  \gamma^I_{ p \dot q} S^{\dot q} \ .
\end{equation}

iii) $N_L =0, N_L =2, 3, \dots, n$ solutions correspond to the
sigma models with
$(4,1)$ world-sheet supersymmetry; the GS formulation of the heterotic string
theory with manifest space-time supersymmetry is equivalent to the NSR form
with the world-sheet supersymmetry. However, for these backgrounds only the
right-handed space-time supersymmetry is available. All left-handed
supersymmetries are broken.
Therefore only the right-moving world-sheet spinor can be converted into the
space-time  spinor  $SO(8)$ vector
\begin{equation}
\lambda_{+}^I = \alpha^{\dot p}  \gamma^I_{\dot p q} S^q \ .
\end{equation}
All solutions in the group $ N_L =2, 3, \dots, n$ are described by the
non-singular bottomless throat geometry.

The ${\bf M}^8$ monopoles seem to provide the best laboratory for the
exploration of
the space-time supersymmetry versus world-sheet supersymmetry.

\section{Exact    F \& H monopoles on $M^4$}
In this section we are going to discuss special examples of our
solution for the case that the spatial part is 4-dimensional, i.e.\
if we have only one non-trivial internal direction, $x^4$.  Most of
the relevant features of F \& H monopoles are already present in this
case. Similar to the $M^8$ for special configurations we can
here expect an enhancement of the world sheet supersymmetry.

We start with general 5-dimensional solution and then we will
consider special examples. For that it is reasonable to rotate
our harmonic functions into a new basis by performing the
$O(1,1)$ transformation
\begin{equation}
\left( \begin{array}{c}  \chi^{(1)}\\ \chi^{(2)} \end{array} \right)
= \frac{1}{\sqrt{2}} \left( \begin{array}{cc}  1 & 1 \\-1 & 1
\end{array} \right) \left( \begin{array}{c}  \chi^{L}\\ \chi^{R}
\end{array} \right)
\end{equation}
The metric and the dilaton are then given by ($y= x^5, \dots x^9$)
 \begin{equation} \label{5dsol}
ds^2 = - dt^2 + d\vec y^2 + 4 \chi^{(1)} \chi^{(2)} d {x^i}^2 +
   \frac{\chi^{(2)}}{\chi^{(1)}} \left(dx^4 +{A}^{(1)}_i  d x^i \right)^2
\qquad
e^{4 \hat{\phi}} = 4 (\chi^{(2)})^2 \ .
\end{equation}
The non-diagonal term in the metric is defined as
\begin{equation} \label{5dF}
F = F^{(1)}_{ij} \equiv \partial_i A^{(1)}_j - \partial_j A^{(1)}_i =
2 \epsilon_{ijm} \partial_m \chi^{(1)}
\end{equation}
The nontrivial part of the 3-form field defines the second gauge field
\begin{equation}  \label{5dH}
H = F^{(2)}_{ij}= \partial_i \hat B _{4j} - \partial_j \hat B_{4i }=
2 \epsilon_{ijm} \partial_m \chi^{(2)}
\end{equation}
We would like to understand the T duality properties of the  uplifted F \& H
monopoles. For this purpose we will perform  Buscher \cite{Bush}
transformation over the solution. The result is very simple: one has 
to change
$\chi^{(1)}$ into $\chi^{(2)}$ and back:
\begin{equation}
\chi^{(1)} \quad  \Longleftarrow T  \Longrightarrow    \quad    \chi^{(2)}
\end{equation}
i.e.\ the compactification radius $g_{44}$ is inverted
and the two gauge fields  F$ \& $H  are exchanged.

The Yang-Mills fields which have to be added to the solutions for avoiding
$\alpha'$ corrections to supersymmetry  parameterize an $SO(4)$. We
will have the space-time indices in a lower position and the Yang-Mills one
in the upper position. In the tangent space there are no zero components of
the vector field, which means also that there are no time components in the
curved space.
$$V_0{}^{ij}
=  V_0{}^{i4} = V_t{}^{ij}=V_t{}^{ij}=0$$
The vector components are
\begin{equation}
V_l{}^{ij} = e^{2U} \delta_{l\, [i} \partial _j\left  ( \ln \chi^{(1)}
\chi^{(2)}\right ) \qquad  V_l{}^{i4} = {1\over 2} e^{2U} \epsilon_{lim}
\partial _m \left ( \ln {\chi^{(1)}\over
\chi^{(2)}}\right )
\label{YM1}\end{equation}
The  fourth components of the vector fields which become scalars
in four dimensions are
\begin{equation}
V_4{}^{ij} \equiv  \Phi^{ij} = -{1\over 2}  e^{2U} \epsilon_{ijm} \partial
_m\left  ( \ln \chi^{(1)}
\chi^{(2)}\right ) \qquad  V_4{}^{i4} \equiv \Phi^{i4} = {1\over 2} e^{2U}
\partial _i \left ( \ln {\chi^{(1)}\over
\chi^{(2)}}\right )
\label{YM2}\end{equation}
Under T duality we have again to exchange $\chi^{(1)}$ and $\chi^{(2)}$
which means that only $V_l{}^{i4}$ and $ \Phi^{i4}$ change the sign.
\pagebreak

\noindent
Let us establish the connection with the previously known
exact heterotic monopoles \footnote{The $M^4$ magnetic solution without
the non-Abelian part was presented recently in \cite{CvT}.}.

\bigskip

\noindent
{\em A. H monopoles ($N_L =1$)} \nopagebreak

\noindent
For this example the F field (\ref{5dF}) is absent,
\begin{equation}
F=0 \ .
\end{equation}
We choose for the harmonic functions
\begin{equation}
\chi^{(1)} =1/2 \qquad , \qquad
\chi^{(2)}=1/2 + 1/ \sqrt 2 \, \sum _{s} {P_s \over |\vec x - \vec x_s|} \ .
\end{equation}
We have taken here the general ansatz as a multi center solution. This,
however, is only consistent if we restrict the positions or the
charges at the centers. We will come back to this point at the end of this
section.
Our solution on ${\bf  M}^5 $ becomes then
 \begin{equation}
\begin{array}{l}
\hat{ds}^2 = - dt^2  +  {\cal V}^{-2}  (dx^i dx^i +  dx^4 dx^4)\ ,  
\qquad  F=0
\\
\nonumber\\
 e^{2 \hat{\phi}}=  2\,  \chi^{(2)}  = {\cal V}^{-2} \quad , \quad
\partial_i \hat B_{4j } - \partial_j \hat B_{4i } = \epsilon _{ijm} \partial_m
e^{2 \hat{\phi}} \end{array}
\end{equation}
and only ${\bf  M}^4 $ is non-trivial. The Yang-Mills fields are
\begin{equation}
V_l{}^{ij} = - 2 \delta_{l\, [i} \partial _{j]}  {\cal V}
 \qquad  V_l{}^{i4} =   \epsilon_{lim} \partial _m   {\cal V} \end{equation}
\begin{equation}
  \Phi^{ij} =    \epsilon_{ijm} \partial _m {\cal V}  \qquad   \Phi^{i4} =
\partial _i  {\cal V}\end{equation}
This solution with $N_L=1$ has self-dual Yang-Mills fields as different from
the general case (\ref{YM1}), (\ref{YM2}).
\begin{equation}
  V_l{}^{i4} =   {1\over 2} \epsilon_{ikm} V_l{}^{km} \qquad , \qquad
 \Phi^{i4} = {1\over 2} \epsilon_{ikm} \Phi {}^{km}
\end{equation}
This self-duality is the source of the enhancement of the left-handed
supersymmetry, since the integrability condition for the Killing spinors is
available. This allows to promote this $N_L=1$ solution to a supersymmetric
solution of the type II string with one half of supersymmetry unbroken. It
results also in (4,4) world-sheet supersymmetry for the corresponding sigma
model.

\bigskip

\noindent
{\em B. F monopoles ($N_L =1$)}

\noindent
This is another example with the same left-handed oscillation number.
Now, the H field (\ref{5dH}) is absent
\begin{equation}
H=0 \ .
\end{equation}
We take the harmonic functions
\begin{equation}
\chi^{(2)} =1/2 \qquad , \qquad \chi^{(1)}=\epsilon/2 + 1/ \sqrt 2
\, \sum _{s} {P_s \over |\vec x - \vec x_s|}
\end{equation}
Now the 3-form and the dilaton is absent and the non-trivial
metric on ${\bf  M}^4 $ is the self-dual multi center metric
\begin{eqnarray}
\hat{ds}^2 = - dt^2  + {\cal V}^{-2} dx^i dx^i +  {\cal V}^{2} (dx^{4} +
\omega_i dx^i)^2, \qquad H=0, \qquad  e^{2 \hat{\phi}}= 1
\end{eqnarray}
where
\begin{equation}
{\cal V}^{-2} \equiv {\chi^{(1)} \over \chi^{(2)}} = \epsilon +  \sum _{s} {
\sqrt 2 P_s \over |\vec x - \vec x_s|}
\end{equation}
\begin{equation}
{\vec \nabla}  ({\cal V}^{-2} ) = {\vec \nabla}\times{\vec\omega}
\end{equation}
This is the multi center Gibbons-Hawking metric \cite{gi/ha}.
Special cases are:
$\epsilon=0, s=1$ (one center) which is the flat Minkowski space,
$\epsilon=0, s=2$ (two center) is the Eguchi--Hanson
instanton and for $\epsilon=1$  this metric corresponds to the
multi-Taub-NUT spaces. Again we have certain restriction for
the charges or positions of the center (see below).

Under Buscher duality transformations \cite{Bush} F monopoles are
transformed into the H monopoles and back. This is obvious from the fact
that under this duality transformation both gauge fields are exchanged.

The F  monopoles Yang-Mills fields are
\begin{equation}
V_l{}^{ij} = - 2 \delta_{l\, [i} \partial _{j]} {\cal V}
 \qquad  V_l{}^{i4} =  - \epsilon_{lim} \partial _m {\cal V} \end{equation}
\begin{equation}
  \Phi^{ij} =    \epsilon_{ijm} \partial _m{\cal V}  \qquad   \Phi^{i4} = -
\partial _i {\cal V} \end{equation}
This solution with $N_L=1$ has anti-self-dual Yang-Mills fields as different
from the general case (\ref{YM1}), (\ref{YM2}).
\begin{equation}
  V_l{}^{i4} = -  {1\over 2} \epsilon_{ikm} V_l{}^{km} \qquad , \qquad
 \Phi^{i4} =  -{1\over 2} \epsilon_{ikm} \Phi {}^{km}
\end{equation}

Here again we are dealing with enhanced supersymmetries. The solution has
one half of unbroken supersymmetries in the type II string and on the
world-sheet we have a (4,4) supersymmetric sigma model.

\bigskip

\noindent
{\em C. T self dual monopoles ($N_L \geq 2$)}

\noindent
We know that under T-duality both gauge fields get exchanged. Therefore
to have a self dual solution we assume that both gauge fields
(\ref{5dF}) and (\ref{5dH}) are equal ($\chi^{(1)} = \chi^{(2)}$)
\begin{equation}
F=H \ .
\end{equation}
It corresponds to uplifted $a=1$ extreme massive magnetic black holes
\cite{BKO}.  In notation of this paper they have $\chi^L =0$ and
\begin{equation}
\chi^{(1)} = \chi^{(2)}=1/2 + 1/ \sqrt 2 \, \sum _{s} {P_s \over
|\vec x - \vec x_s|}
\equiv \chi
\end{equation}
The non-trivial metric on ${\bf  M}^4 $ and the dilaton are
\begin{equation}
ds^2 = - dt^2 +  e^{4 \hat{\phi}} d {x^i}^2 +
  \left(dx^4 +{A}^{(1)}_i  d x^i \right)^2 \qquad
e^{4 \hat{\phi}} = 4 ( \chi ) ^2 \ .
\end{equation}
The Yang-Mills part of the solution is
\begin{equation}
V_l{}^{ij} = -2  \delta_{l\, [i} \partial _{j]}e^{-2  \hat{\phi}}
 \qquad  V_l{}^{i4} =0
\label{YMv}\end{equation}
The  fourth components of the vector fields which become scalars
in four dimensions are
\begin{equation}
 \Phi^{ij} = \epsilon_{ijm} \partial _me^{-2  \hat{\phi}}
 \qquad   \Phi^{i4} = 0\label{YMs}\end{equation}
This is an $SU(2)$ non-Abelian field in agreement with \cite{KO}.
The Yang-Mills field here is not self-dual. Therefore this solution can be
embedded into type II string only as a solution with one quarter of
supersymmetry unbroken, since all left-handed supersymmetries are broken. This
leads to (4,1) supersymmetry on the world sheet, i.e. the enhancement of
supersymmetries takes place only in the right-handed sector of the theory.

\bigskip

\noindent
{\em  D. Massless monopoles ($N_L =0$)}

\noindent
These solutions are given by $\chi^R = const$, which
means that both gauge fields (\ref{5dF}) and (\ref{5dH})
 differ only by a sign
\begin{equation}
F= - H \ .
\end{equation}
Thus, we define our harmonic functions as
\begin{equation} \label{first}
\chi^{(1)} =1/2 + 1/ \sqrt 2 \, \sum _{s} {(P_{\rm vec})_s
\over |\vec x - \vec x_s|} \quad , \quad
\chi^{(2)}=1/2 - 1/ \sqrt 2 \, \sum _{s} {(P_{\rm vec})_s
\over |\vec x - \vec x_s|}
\end{equation}
The  metric and the dilaton are now
\begin{equation}
ds^2 = - dt^2  +4 \chi^{(1)} \chi^{(2)} d {x^i}^2 +
   \frac{\chi^{(2)}}{\chi^{(1)}} \left(dx^4 +{A}^{(1)}_i  d x^i \right)^2
\qquad
e^{4 \hat{\phi}} = 4 (\chi^{(2)})^2 \ .
\end{equation}
This solution describes in four dimensions a massless monopole that
was uplifted into the 5-dimensional stringy geometry.

Under  T duality transformation  the massless uplifted monopoles are not
invariant since $\chi^{(1)} \neq  \chi^{(2)}$.  If  we would perform  Buscher
\cite{Bush}  transformation over the solution and combine it with the charge
conjugation \footnote{In what follows we will use a notation $T_d$ for T
duality since in the context of charge conjugation $C$ one may expect the
symbol $T$ to be associated with time reflection.}
\begin{eqnarray}
\chi^{(1)} \quad  \Longleftarrow &T_d & \Longrightarrow    \quad    \chi^{(2)}
\\
(P_{\rm vec})_s  \; \Longleftarrow  &C & \Longrightarrow \;  - (P_{\rm vec})_s
\end{eqnarray}
we would find that the solution is invariant. Indeed, for the massless
monopoles
\begin{eqnarray}
\chi^{(1)} \quad  \Longleftarrow &T_d \times C & \Longrightarrow    \quad
\chi^{(1)} \\
\chi^{(2)}   \quad \Longleftarrow  &T_d \times C & \Longrightarrow  \quad
\chi^{(2)}
\end{eqnarray}
This property of the massless monopoles was not predicted before and the fact
that it involves T duality and changing the monopole into the anti-monopole
$T_d \times C$ does not seem to follow from any known principles. It is
observed here as a property of the explicit solution.

The YM fields fields for the massless monopoles are given  in 
eqs. (\ref{YMv}),
(\ref{YMs}) with the harmonic functions  defined in eq.\ (\ref{first}).
There is no special simplifications: the non-Abelian fields
belong to the $SO(4)$ gauge group. This solution breaks one half of the
supersymmetry of the heterotic string,  and
corresponds to (4,1) supersymmetric sigma model.

\bigskip

\noindent
{\em E. The moduli matrix ${\cal M}$}

\noindent
Thus we have listed here four special type of F  \& H monopoles. 
Two of them, H
monopoles and F  monopoles have the four-dimensional projection in which the
metric and the dilaton are  the same for both solutions and coincide 
with those
of the extreme magnetic $a=\sqrt 3$ black holes
\begin{equation}
ds^2_{\rm str}= -  dt^2 + e^{-4 U} d\vec{x}^2\ ,    e^{-4U} = \chi
                                              = e^{4\phi} =1+ {2M \over r}\ ,
\qquad M=P/\sqrt 2
\end{equation}
The difference comes in moduli ${\cal M}$ and in the vector fields since the
right-handed harmonic function $\chi^R$ is the same for both solutions but the
left-handed one differs by the sign of the charge.
For our  case here the moduli metric ${\cal M}$ is given
by
\begin{equation}
{\cal M} =    \pmatrix{ {\chi^{(1)} \over \chi^{(2)}} & 0 \cr
\cr
0   & {\chi^{(2)} \over \chi^{(1)}}  \cr}
\end{equation}

For the H monopole we have therefore
for the non-vanishing component of moduli and vector fields
\begin{equation}
{\cal M}_H =    \pmatrix{ \frac{1}{1 + \frac{2 M}{r}}  & 0  \cr
\cr
0   & 1 + \frac{2 M}{r}\cr}\ , \qquad \left (\matrix{
F^{(L)}_{ij}\cr
F^{(R)}_{ij}\cr
}\right )= \epsilon_{ijm}\partial_m  {P\over  r} \left (\matrix{
-1\cr
\cr
+1\cr
}\right )
\end{equation}
For the F monopoles the non-vanishing components of moduli and vector fields
are
\begin{equation}
{\cal M}_F = {\cal M}_H^{-1} = \pmatrix{ 1 + \frac{2 M}{r} & 0  \cr
\cr
0   &  \frac{1}{1 + \frac{2 M}{r}}  \cr} \ , \qquad \left (\matrix{
F^{(L)}_{ij}\cr
F^{(R)}_{ij}\cr
}\right )= \epsilon_{ijm}\partial_m  {P\over  r} \left (\matrix{
+1\cr
\cr
+1\cr
}\right )
\end{equation}
The T self dual solution in four dimensional form becomes equal to the extreme
$a=1$ dilaton black hole:
\begin{equation}
ds^2_{\rm str}= -  dt^2 + e^{-4 U} d\vec{x}^2\ ,   \qquad  e^{-4U} = \chi
                                              = e^{4\phi} =(1+ {2M \over r})^2
\ ,  \qquad M=P/\sqrt 2
\end{equation}
The moduli fields are constant and  vector fields are only the right-handed
ones.
\begin{equation}
{\cal M} =  {\bf 1}_{12} \ ,    \qquad \left (\matrix{
F^{(L)}_{ij}\cr
F^{(R)}_{ij}\cr
}\right )= \epsilon_{ijm}\partial_m  {P\over r} \left (\matrix{
0\cr
\cr
+1\cr
}\right )
\end{equation}
Finally, the massless monopoles in the four-dimensional world are the ones
whose metric and the dilaton are:
\begin{equation}
ds^2_{\rm str}= -  dt^2 + e^{-4 U} d\vec{x}^2\ ,   \qquad  e^{-4U}
                 = e^{4\phi} =1 -  {1 \over 2}
\left({P_{\rm vec} \over r}\right)^2 \ ,  \qquad M=0
\end{equation}
There are some non-vanishing components of the moduli fields  and and the
vector fields are only the left-handed ones.
\begin{equation}
{\cal M}  = \pmatrix{ \frac{ r + 2 M}{ r - 2 M} & 0\cr
0 & \frac{ r - 2 M}{ r + 2 M} \cr
}\ ,    \qquad \left (\matrix{
F^{(L)}_{ij}\cr
F^{(R)}_{ij}\cr
}\right )= \epsilon_{ijm}\partial_m  {P_{\rm vec}  \over  r} \left (\matrix{
+1\cr
\cr
0\cr
}\right )
\end{equation}

\bigskip

\noindent
{\em F. Remark about the multi center case}

\noindent
All our magnetic (Abelian) gauge fields are given by Dirac monopole
solutions. As a consequence in order to formulate the solution consistently
one has to remove  the so-called Dirac-singularities.
This is in principle not difficult, one has only to take
care that it can be done for all centers simultaneously. We are describing
this procedure in detail in the Appendix B. As result we find that for
all F \& H monopoles  i) the position of the centers can be arbitrary,
but all charges may differ only by the sign and for the other cases ii) all
centers have to line-up. For the second case the charges are arbitrary. By
fixing the gauge one could also relax the line-up restriction, but not in a
gauge invariant way. The two-center solutions can always be placed on a
line, therefore there are no restrictions. The subtlety is relevant
starting from three centers.

\vskip 0.6 cm
As we have already pointed out above, we have listed here four particular
special cases of F \& H monopoles. The singularities of these solutions have
been studied before. The $N_L=0$ massless monopoles and $N_L=1$ F \& H
monopoles are singular in four-dimensional geometry.  The majority of
solutions, i.e. the generic
$N_L=2,3,4, \dots$  solutions have an arbitrary  non-vanishing
values of F and H and are non-singular bottomless holes from the point of view
of the  four dimensional stringy geometry.

\section{Discussion}
The large family of monopole solutions described in this paper provides an
interesting
area of study of non-perturbative effects in supersymmetric gravity. The fact
that these configurations have a non-trivial Euclidean geometry on an eight
dimensional manifold ${\bf M}^8$ has made these new solutions a most
interesting objects realizing the relations between unbroken space-time
supersymmetry and world-sheet supersymmetry. The role of $SO(8)$ gauge group
with its particularly remarkable relations between vector and spinor
representations acquires a new and beautiful aspect when applied to
supersymmetric stringy monopoles. The unique property of such monopoles to
admit complex structures in the corresponding sigma models  is related
immediately to the fact  that the space-time Killing spinors for magnetic
configurations are constant in the stringy frame. Therefore we have observed
the enhancement of world-sheet supersymmetries for all stringy monopoles which
live on
the Euclidean ${\bf M}^8$ manifold. These monopoles, although discovered as
a soliton type solutions of the classical field equations of 
supergravity seem
to have the most stable properties concerning the quantum corrections. The
supersymmetry in the space-time as well as the supersymmetry at the 
world-sheet
are free of anomalies. The supersymmetric non-renormalization theorems do not
have any obvious obstructions and therefore the ${\bf M}^8$ monopoles give us
an example of
rather reliable non-perturbative BPS states of the superstring theory.

\section*{Acknowledgements}
The work of K.B. is supported by the DFG and by a grant of
the DAAD. He would also like to thank the Physics Department
of the Stanford University for its hospitality.
The work of  R.K. is  supported by  the NSF grant PHY-8612280.

\newpage

\section {Appendix A: Spin connections}

Now we define the vielbeins. The zehnbein can be written as
\begin{equation}
\hat{E}^A{}_{M} = \left( \begin{array}{cc} \matrix{
e^0{}_t & 0 \cr
0 &  e^i{}_{\underline i}
\cr
} &
0 \\  E^{a}{}_{{\beta}} A^{(1){\beta}}{}_{\underline i}
  & E^{a}{}_{\alpha} \end{array} \right)
\end{equation}
The curved three dimensional indices  are underlined.  The same can be
rewritten as
\begin{equation}
\hat E^A = \hat{E}^A{}_{M} dX^M = \{ e^0=  dt \ , e^i= e^{i}{}_{\underline  i}
dx^ {\underline  i} \ , \;
E^a= E^{a}{}_{{\beta}} A^{(1){\beta}}{}_{\underline i} dx^{\underline i} +
E^{a}{}_{\alpha}dx^\alpha \}
\end{equation}
The corresponding vierbeine are given by
\begin{equation}
e^{0}{}_{t} = 1 \qquad , \qquad e^{i}_{\ \underline{j}}=
e^{-2U} \delta^i_{\ j}\\
\end{equation}
The six-dimensional vielbein is defined in terms of our 12 harmonic functions
as
\begin{equation}
E^{a}{}_{\alpha} = \delta^{a \beta } \left[ \delta_{\beta \alpha}  -
\frac{1}{|\vec{\chi}^R|^2 + (\vec{\chi}^R \vec{\chi}^L)} \left(
(1-\sqrt{1 - \frac{|\vec{\chi}^L|^2}{|\vec{\chi}^R|^2}}) \chi^R_{\beta}
\chi^R_{\alpha} \ +\  \chi^L_{\beta} \chi^R_{\alpha} \right)\right]
\end{equation}
The inverse quantities
are: $$G^{\alpha\beta} = \delta^{\alpha\beta} + 2 e^{4U} [ \chi^L_{\alpha}
\chi^L_{\beta} + \frac{|\vec{\chi}^L|^2}{|\vec{\chi}^R|^2}
\chi^R_{\alpha} \chi^R_{\beta} + 2 \chi^L_{(\alpha} \chi^R_{\beta)}] $$
$$ E_{\ a}^{{\alpha}} = \left [\delta_{ \alpha  \beta}  +
\frac{\sqrt{2} e^{2U}}{|\vec{\chi}^R|} \left( (1 - \sqrt{1 -
\frac{|\vec{\chi}^L|^2}{|\vec{\chi}^R|^2}}) \chi^R_{\alpha}
\chi^R_{\beta} \ -\  \chi^R_{\alpha} \chi^L_{\beta} \right)\right]
\delta^\beta{}_a$$
The results for the metric spin connection one-form consist of
three type of terms representing the $SO(3)$ rotation  $M^{[ij]}$ , 
the $SO(6)$
rotation $M^{[ab]}$ and the non-diagonal terms  $M^{[ai]}$.
\begin{eqnarray}
W _{ij}&=&2 (\partial_{[i }e^{2U})  e_{j]} -{1\over 2}  F_{ij,  a}  E^a\\
\nonumber\\
W_{ab}&=& E_{[a}{}^\alpha \delta_i{}^ { \underline i} 
( \partial_  { \underline
i}    E_{\alpha b]}) e^i \\
\nonumber\\
W_{ai}&=& F_{ik}{}^\alpha E_{\alpha a} e^k - {1\over 2} E_a{}^{\alpha} e^{2U}
\delta_i{}^ { \underline i} ( \partial_  { \underline i}  G_{\alpha \beta})
E^\beta{}_c E^c
\end{eqnarray}
All components of spin connection related to the boosts are vanishing:
\begin{equation}
W_{0A} =0
\end{equation}

To calculate the tangent space three form $H_{ABC}$ we will use the 
expression
for the curved space three form and the zehnbeins
\begin{equation}
H_{ABC}= E_A{}^M E_B{}^N E_C{}^L H_{MNL}
\end{equation}
where $H_{MNL} = 3  \partial_{[M}  B_{NL]} $ and the components of the 
two-form
$B_{MN}$ are defined in eq. (\ref{antisym})

For all monopoles  $H_{tNL}=0$.
It follows that
\begin{equation}
H_{0BC}=0 \qquad  H_{0B}\equiv H_{0BC} E^C =0
\end{equation}
We will get 3 types of terms for $H_{ABC} E^C$.  Since none of the indices in
$H_{ABC} E^C$  takes the value $0$ we could rewrite it as $H_{IJK} E^K$. We
will get 3 types of terms for it.
\begin{eqnarray}
H_{ij}&=& H_{ijk} e^k + H_{ija} E^a\\
\nonumber\\
H_{ab}& =&H_{abk} e^k + H_{abc} E^c\\
\nonumber\\
H_{ai} &=& H_{aik} e^k + H_{aic} E^c
\end{eqnarray}
For the one center solution only the term $H_{ijk}$ vanishes, for the multi
center solution the situation is more complicated. Thus in general all
components of  $H_{IJ}$ are non-vanishing.

There are two combinations of spin connection and three form which we need to
know
\begin{equation}
\Omega_{\pm IJ} = W_{IJ} \pm H_{IJ}
\end{equation}
They are both taking values in $SO(9)$ group, in general. For some particular
solution they may take values in much smaller groups which are subgroups of
$SO(9)$  but they do not need more than
$SO(9)$ and since there are no boost components in neither metric spin
connections nor in the three forms, this concerns both generalized spin
connections $\Omega_{\pm IJ}$.

\section{Appendix B: Multi center solutions}

Here we are going to discuss possible restrictions for parameters  of
the multi center solution.  We consider for simplicity the ${\bf M}^4$
monopoles presented in Sec. 5 . The solution is defined in terms of two
harmonic functions.  To describe a multi center solution we take these
harmonic functions in the form
\begin{equation}   \label{chi}
\chi^{(1)} = a^{(1)} + \sum_{s=1}^{k+1} \frac{p^{F}_s}{|\vec{x} -
\vec{x}_s|} \quad , \quad \chi^{(2)} = a^{(2)} + \sum_{s=1}^{k+1}
\frac{p^{H}_s}{|\vec{x} - \vec{x}_s|}
\end{equation}
where $\vec{x}_s$ are the positions and $p_s$ are the charges
of the centers.
For the gauge fields in the multi-center case we can also make an
ansatz as a sum over different one-center solutions
\begin{equation}
\vec{A}^{(1/2)} = \sum_{s=1}^{k+1} \vec{A}_s^{(1/2)} \ .
\end{equation}
We are describing here Dirac monopole solutions which are
globally not defined. To remove the  Dirac singularities
one has to introduce for every center two different coordinate
patches. In  each of them one defines the gauge field without
a singularity and finally one has to glue together all patches.
For one center the different gauge fields are then given by
\cite{eg/gi/ha}
\begin{equation}  \label{solA}
\vec{A}^{(1/2)}_s = \frac{ p_s^{(F/H)}}{2\, r_s(z-z_s \pm r_s)}
 \left( (y-y_s)\,dx - (x-x_s)\, dy \right)
 = \frac{1}{2}\, p_s^{(F/H)} ( \mp 1 + \cos \theta_s) d \phi_s
\end{equation}
where $r_s^2 = (x-x_s)^2 + (y-y_s)^2 + (z-z_s)^2$ and $\theta_s$ and
$\phi_s$ are the angular variables of the center $s$. For every gauge field
we have a ``+'' and a ``-'' part which are singular for $\theta_s = 0$ or
$\pi$ and we have to take the non-singular one in the different patches. In
the overlapping region both parts are connected by a gauge transformation
which is equivalent to a shift in the $x^4$ coordinates (see (\ref{5dsol})
and also \cite{gi/ha})
\begin{equation}
x^4_{(s)} \rightarrow x^4_{(s)} \mp p_s^{F} \phi_s \ .
\end{equation}
(the ``$\mp$'' ambiguity comes from the fact that one can approach the
overlapping region from two different patches).  Since we have to
identify the field configuration in the overlapping region we find that
$x^4_s$ has to be periodic with the period of $2 \pi p_s^{F}$ which
means that the $x^4$ direction is a circle with the radius $p_s^{F}$.
This has to be done for every center.  This procedure can be
done only if the compactification radius of $x^4$ is the same for all
centers. Otherwise one could not put together all the the different
coordinate patches for all centers.  But since the period or radius of
the $x^4$ differs if the magnetic charges differs from center to center
we get the result that all magnetic charges are the same up to a sign
or zero, i.e.
\begin{equation}
p_s^{F} =  p^{F} \, \eta_s^{F} \ , \ \  \eta_s^{F} = 0, \pm 1
\end{equation}
There  is a second possibility to remove all Dirac singularities
consistently. Namely if all centers are on a line, e.g.\ if $x_s$ and
$y_s$ are equal all centers line-up parallel to the $z$-direction. Then
we can introduce for all centers the same coordinate system with the
same angular variable $\phi$, $\vec{A}^{(1)}d\vec{x} \equiv
A_{\phi}^{(1)} d\phi$.  Consequently, the periodicity condition for $x^4$
is now given by
\begin{equation}
x^4 \simeq x^4 + P^{F} \phi
\end{equation}
where $P^{F} = \sum_{s=1}^{k+1} p_s^{F}$, the total magnetic
charge. Again the $x^4$ direction is compactified on a circle with
radius proportional to the total magnetic charge.

These results were related to the fact that  $\vec{A}^{(1)}$  was the
KK gauge field in the metric and thus restricts the $p_s^{F}$. A simple $T$
duality, however, exchange both gauge fields  $\vec{A}^{(1)}$ and
$\vec{A}^{(2)}$  and thus yields the same restrictions for $p_s^{H}$ as
well.

To summarize, in order not to have Dirac singularities in the ${\bf
  M}^4$ metric we have two possibilities i) all centers have up to the
sign the same magnetic charges (or zero) $p^{F}$ and $p^{H}$, or ii) all
centers have to line-up. For the harmonic functions for the multi
center solutions this means
\begin{equation}
\begin{array}{l}  \label{harm}
i) \qquad \chi^{(1)} = a^{(1)} + p^{F} \sum_{s=1}^{k+1}
 \frac{\eta_s^{F}}{|\vec{x} - \vec{x}_s|} \quad , \quad \chi^{(2)} =
 a^{(2)} + p^{H}\sum_{s=1}^{k+1}
 \frac{\eta_s^{H}}{|\vec{x} - \vec{x}_s|} \\
ii) \qquad
  \chi^{(1)} = a^{(1)} + \sum_{s=1}^{k+1} \frac{p^{F}_s}{|\vec{x} -
 z_s|} \quad , \quad \chi^{(2)} = a^{(2)} + \sum_{s=1}^{k+1}
 \frac{p^{H}_s}{|\vec{x} - z_s|}
\end{array}
\end{equation}
where $\eta_s = 0, \pm 1$. In both cases the $x^4$ coordinate has to be
compact. Obviously, the two-center case falls into the second
possibility. The first non-trivial case are three centers.

\vspace{3mm}

Now we turn to the question what do these restrictions mean for the
torsion or antisymmetric tensor. We start with the calculations of
the Chern-Simons term, which is part of the torsion in $D=4$.
Using (\ref{chi}) and (\ref{solA}) we find
\begin{equation}
\begin{array}{l}
(A_i^{(1)} F_{jl}^{(2)} + A_i^{(2)} F_{jl}^{(1)} + cycl.perm.) \sim \\[1mm]
\sim\, \epsilon_{ijl} \left( A^{(1)}_m \partial_m \chi^{(2)} +
 A^{(1)}_m \partial_m \chi^{(2)} \right) = \\[1mm]
=\, \epsilon_{ijl} \sum_{st} \frac{p_s^{F} p_t^{H} + p_s^{H} p_t^{F}}{
  r_s(z-z_s \pm r_s) \; r_t^3} \left[ (x-x_s)(y-y_t) - (y-y_s)(x-x_t)
\right]\ .
\end{array}
\end{equation}
Since our $D=4$ torsion has to vanish there are two
possibilities. Either the Chern-Simons term and the antisymmetric
tensor vanish each or they cancel against each other
\footnote{Note, the 4-dimensional torsion is given by \cite{MS}
$H_{\mu\nu\rho} = \partial_{\mu} B_{\nu\rho} - 2 (A_{\mu}^{(1)}
F_{\nu\rho}^{(2)} + A_{\mu}^{(2)}F_{\nu\rho}^{(1)}) + cycl. perm.
$, with $B_{\mu\nu} = \hat{B}_{\mu\nu} + 2(A_{\mu}^{(1)} A_{\nu}^{(2)} -
  A_{\nu}^{(1)} A_{\mu}^{(2)} ) $
}. The Chern-Simons
term vanishes under the two conditions:
\begin{equation}
\begin{array}{ll}
i) & p_{s}^{F} p_{t}^{H} = - p_{t}^{F} p_{s}^{H} \\
ii)& \mbox{all centers line-up, i.e. }\  x_s=x_t \, , \, y_s=y_t\;
  \forall \ s,t \ .
\end{array}
\end{equation}
The first condition means that either $p_{s}^{F}$ or $p_{s}^{H}$
is identical zero for all centers, i.e.\ only F or H monopoles.
The line-up condition $ii)$ coincides
with the condition $ii)$ in (\ref{harm}). If these conditions are
fulfilled we find  that the $D=4$ antisymmetric tensor has
to be zero, too. Also the $D=5$ antisymmetric tensor (see footnote)
is zero,
since in the line-up case the gauge fields have only one
unique $\phi$-component and in the other case one gauge field
vanishes. Thus,
\begin{equation}
\hat{B}_{ij} = B_{ij} =0 \quad , \  i,j = 1,2,3
\end{equation}
 Finally, we have to investigate the possibility that
the antisymmetric tensor part
cancels the Chern-Simons part. This is in our case
possible since for our gauge fields $F^{(1)}\wedge F^{(2)} =0$.
\begin{equation}
\begin{array}{rcl}
3 \partial_{[m} B_{np]} &=& 6 (A_{[m}^{(1)} F_{np]}^{(2)} +
 A_{[m}^{(2)} F_{np]}^{(1)}) \\[1mm]
 &=&  \epsilon_{mnp} (A_{l}^{(1)} \partial_l \chi^{(2)} +
    A_{l}^{(2)} \partial_l \chi^{(1)}) \ .
\end{array}
\end{equation}
This equation has a solution if the gauge fields are in the
Coulomb gauge ($\partial_l A_l^{(1/2)}=0$) and we find
\begin{equation}
B_{np} =  \epsilon_{npl} (A_l^{(1)} \chi^{(2)} + A_l^{(2)} \chi^{(1)})
\end{equation}
and inserting this expression into the $D=5$
antisymmetric tensor (see footnote) yields
\begin{equation}
\hat{B}_{np} =  \epsilon_{npl} (A_l^{(1)} \chi^{(2)} +
   A_l^{(2)} \chi^{(1)}) - 2(A_n^{(1)} A_p^{(2)} - A_p^{(1)} A_n^{(2)}) \ .
\end{equation}
This procedure is possible independently on the positions of the centers,
but it is somehow unsatisfactory since it is gauge dependent (only for
Coulomb gauge). The situation becomes even worse when we remember that the
gauge invariance of KK gauge fields is related to general covariance of the
5-dimensional theory (translations in the fourth coordinate). From this point
of view it seems to be that the multicenter solution is consistent in 4 as
well as embedded in 5 dimensions only i) if one gauge field vanishes (i.e.\
$F$ or $H$ monopoles) or ii) if all centers line-up. In the last case the
charges could be arbitrary whereas in the case of $F$ or $H$ monopoles
it is necessary that all charges may differ
only by a sign (to remove all Dirac singularities).

\end{document}